\documentclass{article}
\usepackage{spconf,amsmath,graphicx,hyperref}
\usepackage{tabularray}
\usepackage{caption}       
\usepackage{adjustbox}     
\usepackage{amssymb}
\usepackage{multirow}   
\usepackage{enumitem}
\usepackage{booktabs}
\usepackage{tabularx}    
\title{E2E-AEC: IMPLEMENTING AN END-TO-END NEURAL NETWORK LEARNING APPROACH FOR ACOUSTIC ECHO CANCELLATION}
\name{Yiheng Jiang, Biao Tian, Haoxu Wang, Shengkui Zhao, Bin Ma, Daren Chen, Xiangang Li}
\address{Tongyi Lab, Alibaba Group, China}
\begin{document}

%
\maketitle
\begin{abstract}
We propose a novel neural network-based end-to-end acoustic echo cancellation (E2E-AEC) method capable of streaming inference, which operates effectively without reliance on traditional linear AEC (LAEC) techniques and time delay estimation. Our approach includes several key strategies: First, we introduce and refine progressive learning to gradually enhance echo suppression. Second, our model employs knowledge transfer by initializing with a pre-trained LAECbased model, harnessing the insights gained from LAEC training. Third, we optimize the attention mechanism with a loss function applied on attention weights to achieve precise time alignment between the reference and microphone signals. Lastly, we incorporate voice activity detection to enhance speech quality and improve echo removal by masking the network output when near-end speech is absent. The effectiveness of our approach is validated through experiments conducted on public datasets.
\end{abstract}
\vspace{-10pt}
\section{Introduction}
\label{sec:intro}

Acoustic echo cancellation (AEC) is a crucial technology in voice interaction systems, addressing the significant challenge of eliminating acoustic feedback to ensure high-quality audio communication [1].

Traditionally, AEC systems relied on digital signal processing methods, typically starting with time delay estimation (TDE) to align microphone and far-end reference signals in the presence of latency. Following this alignment, adaptive filters, such as those based on the normalized least mean square (NLMS) [2] algorithm or its variants, were employed to estimate and cancel echo [3]. These methods, referred to as linear AEC (LAEC), proved effective under stable linear conditions. However, they struggled to perform adequately when echo paths were time-varying or non-linear, leading to either leakage of echo or suppression of the near-end talker [4].

In recent years, hybrid systems that combine traditional methods with neural networks (NN) have been widely explored [5-7]. In such systems, LAEC is typically employed\\[0pt]
for preliminary echo cancellation, with NN subsequently addressing residual echo suppression. In [5], a multi-filter LAEC followed by a gated complex convolutional recurrent neural network is explored for AEC task. Alternatively, MTFAA-Net [7] implements a hybrid AEC system by integrating TDE, LAEC, multi-scale NN modeling, and an attention mechanism, achieving superior performance.

Recently, some efforts have been made to develop end-toend (E2E) NN methods that bypass TDE and LAEC [8-10]. A major challenge for these E2E approaches is the absence of TDE algorithms, which may lead to severe performance degradation in scenarios with large time delays [11], requiring the model to learn alignment internally. In [8], researchers propose cross-attention as an alternative to TDE and employ a convolutional recurrent network to suppress echo, eliminating the need for LAEC. DeepVQE [9] is another E2E method that employs a distinct attention mechanism to explicitly align microphone and reference signals, while also optimizing other network modules to achieve outstanding performance.

Despite these advancements, hybrid systems remain predominant in industrial applications due to their robustness and reliability. To further advance E2E methods toward realworld deployment, we propose a novel neural network-based E2E-AEC framework. This framework integrates an optimized progressive learning (PL) [12] to jointly tackle noise reduction and echo cancellation. Our model is initialized using a pre-trained LAEC-based NN, enabling effective knowledge transfer from a hybrid AEC system. In addition, we employ an improved attention mechanism guided by loss functions for precise time alignment. Finally, we utilize a loss function for near-end speech voice activity detection (VAD), which could enhance speech quality while enabling strategic masking of far-end echo during near-end silence. Overall, this research highlights the potential of E2E approaches and contributes to advancing this challenging field.

\section{METHODS}
\subsection{Problem Formulation}
In the field of AEC, the microphone signal in time domain is typically modeled as:

\begin{equation*}
y(n)=r(n) * h_{r}(n)+x(n) * h_{x}(n)+v(n) \tag{1}
\end{equation*}
where $*$ denotes the convolution operation, $n$ indexes time samples, $h_{r}(n)$ and $h_{x}(n)$ are acoustic transfer functions [13], $r(n), x(n)$ and $v(n)$ are the far-end reference signal, near-end target speech and additive background noise, respectively. Given observation $y(n)$ and reference $r(n)$, the AEC task focuses on suppressing the echo component $r(n) * h_{r}(n)$. In this work, we employ an E2E-AEC model to not only suppress echo, but also mitigate reverberation and background noise, thereby recovering the anechoic target speech $x(n)$.

\subsection{Overall Network Architecture}
Figure 1 shows the architecture of the proposed E2E-AEC model. The network takes short-time Fourier transform (STFT) features as input and output, with a frame length of 20 ms and a frame shift of 10 ms . Each fundamental recurrent neural network (RNN) [14] block consists of two GRU layers [15], following the TF-GridNet design [16] to capture full-band and sub-band dependencies. The GRU is uni-directional to support streaming inference. The hidden dimension is set to 64 , the kernel size and stride size in the unfold operation are 4 and 1 , respectively.

The input features are encoded by RNN blocks, followed by an attention mechanism that generates aligned ref, with a supervised loss imposed on the attention weights. The aligned ref and mic features are then concatenated and passed through 8 RNN blocks. The network outputs complex convolving masks [9] in the middle and final layers, applied to the mic spectrum to recover the stage-specific targets for PL training. Additionally, a near-end speech VAD prediction is derived from the middle layer. The model has 1.2 M parameters, and as the innovation of this work does not lie in the model structure, our approach can generalize to networks with more or fewer parameters.

\subsection{Progressive Enhancement}
Progressive learning (PL) [12] divides the model into multiple stages, where the training target for each stage has a higher signal-to-noise ratio (SNR) than the previous one. This design ensures that the later stage focuses on reconstructing signals with better quality, guiding the model toward incremental enhancements in speech quality.

\begin{figure}[h]
  \centering
  \centerline{\includegraphics[width=5.5cm]{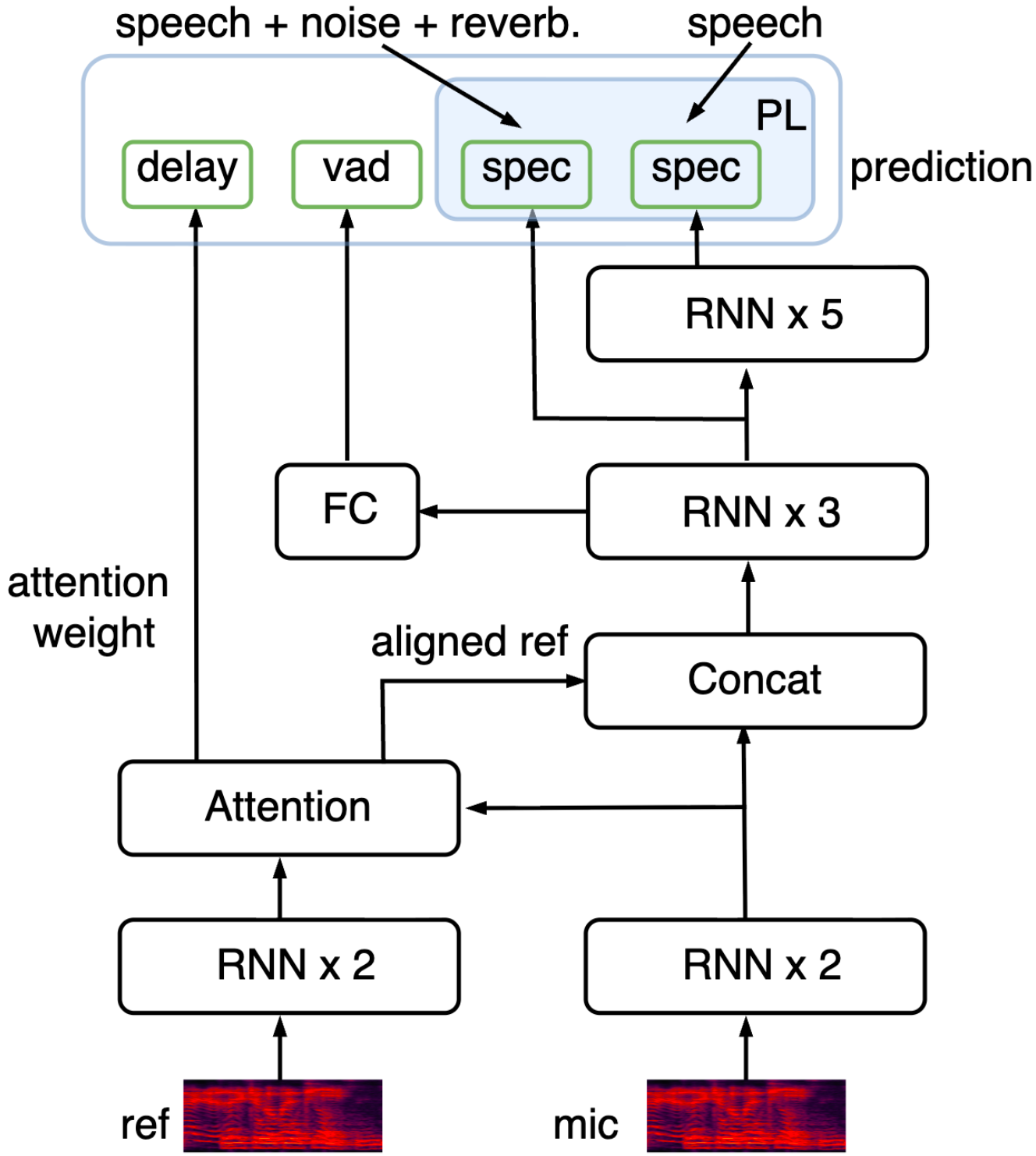}}
\captionsetup{labelformat=empty}
\caption{Figure 1. E2E-AEC system overview. The outputs include: delay (time delay estimation), vad (near-end speech VAD), and spec (spectrum estimations of different PL stages).}
\end{figure}

In AEC scenarios, we modify the PL framework such that stage targets differ by signal components instead of SNR (see Figure 1). The target of the first stage is near-end speech mixed with background noise and reverberation but without echo, which enables the network to focus exclusively on removing dominant echo. The target of the second stage is nearend speech only, refining the output by eliminating residual echo, background noise, and reverberation. By substantially reducing echo in the first stage, the second stage lightens the burden of echo removal, enabling more effective suppression of noise and reverberation. After PL training, the final predicted signal is produced from the last stage.



\subsection{Knowledge Transfer}
Hybrid AEC systems utilize TDE and LAEC for time alignment and initial echo removal, reducing the burden on subsequent modules and improving overall performance [17]. To enable our E2E-AEC model to leverage this, transfer learning can be implemented through two approaches: (1) a teacherstudent framework [18, 19], where a pre-trained hybrid system guides the E2E model, and (2) parameter initialization, which uses the parameters of a pre-trained hybrid model to initialize the E2E-AEC model. Both strategies provide a mechanism for transferring relevant knowledge learned by the hybrid system to the E2E model. In this work, we adopt the second approach for its simplicity and efficiency, leaving the teacher-student approach for future work.

\subsection{Time Alignment}
Inspired by the approach in [9], we introduce an attention mechanism to dynamically align the reference features $R$ and the microphone features $Y$, as illustrated in Figure 1. Both $R$ and $Y$ are assumed to have the shape $\mathbb{R}^{C \times T \times F}$, where $C$, $T$, and $F$ represent the number of channels, time frames, and feature dimensions, respectively.

The time alignment process begins by unfolding $R$ along the time axis, yielding $R_{u} \in \mathbb{R}^{C \times T \times H \times F}$, where $H$ denotes the maximum allowable latency. Correlation between $Y$ and $R_{u}$ is computed via a dot product:

\begin{equation*}
D_{p}(c, t, d)=\sum_{f=0}^{F-1} Y(c, t, f) \cdot R_{u}(c, t, d, f) \tag{2}
\end{equation*}

$D_{p} \in \mathbb{R}^{C \times T \times H}$ is then processed through a convolutional layer to produce a single-channel representation, followed by\\
applying softmax function, resulting in attention weights $A \in \mathbb{R}^{T \times H}$ that represent the probability distribution across different time delays. Finally, the aligned reference features $\tilde{R} \in \mathbb{R}^{C \times T \times F}$ are given by:

\begin{equation*}
\tilde{R}(c, t, f)=\sum_{d=0}^{H-1} A(t, d) \cdot R_{u}(c, t, d, f) \tag{3}
\end{equation*}

This soft alignment method avoids the non-causal problems that may be introduced by traditional TDE relying on absolute delay estimation. More details can be found in [9].

To further enhance the time alignment, we introduce a loss function on attention weights, explicitly encouraging $A$ to perform TDE. The estimated delay is computed as:

\begin{equation*}
D_{e}(t)=\sum_{d=0}^{H-1} A(t, d) \cdot d \tag{4}
\end{equation*}

The target delay is computed using the GCC-PHAT [20] algorithm and discretized into integer classes to represent framelevel delays. Two loss functions are considered in this work: mean squared error (MSE) and cross-entropy loss (CE). The key distinction between them lies in their treatment of prediction granularity: MSE directly penalizes the difference between the predicted delay and the target, encouraging predictions to be as closely as possible to the target. In contrast, CE treats different delays as independent categories, where even neighboring frames are uncorrelated, making it fundamentally different from MSE. As demonstrated later, experiments show that this explicit supervision of attention weights significantly improves alignment performance.

\subsection{VAD Prediction and Masking}
The VAD can enhance the model's ability to differentiate between near-end speech and other signal components in AEC tasks [21]. Following [17], as shown in Figure 1, an intermediate layer employs a fully connected (FC) layer to convert features into frame-level VAD predictions for near-end speech. The binary cross-entropy loss (BCE) is computed between these predictions and the ground truth labels, which are generated using WebRTC-VAD.

To further improve echo suppression during inference, a VAD-masking operation is applied to the output spectrum. The predicted VAD probabilities are first smoothed to stabilize frame-level decisions. For each frame, when the absence of near-end speech is indicated with high confidence (i.e., the probability of non-speech exceeds a predefined threshold), a conditional mask is applied. This mask reduces the magnitude of that frame by a factor in output spectrum, effectively suppressing residual echo during far-end single talk periods.

\subsection{Loss Function}
The loss function for spectrum estimation in PL framework is defined as a weighted combination of modulation loss [22]\\[0pt]
and SNR loss [23], with respective weights of 0.1 and 0.9 . The overall loss to be minimized is expressed as:

\begin{equation*}
L=\lambda_{1} L_{spec1}+\lambda_{2} L_{spec2}+\lambda_{3} L_{delay}+\lambda_{4} L_{vad} \tag{5}
\end{equation*}

Here, $L_{spec1}$ and $L_{spec2}$ represent the spectrum estimation losses in PL framework. $L_{vad}$ is the VAD prediction loss, which is optimized using BCE. $L_{delay}$ refers to the delay prediction loss for time alignment, where the weight $\lambda_{3}$ is set to 100 when using MSE and to 1 when using CE, ensuring that $L_{delay}$ has a similar scale under both loss functions. All other weights are set to 1 .

\begin{table*}[htbp]  
  \centering
  \footnotesize       
  \captionsetup{labelformat=empty}
  \caption{Table 1. The AECMOS and ERLE (measured in dB) results on AEC Challenge blind test sets, including near-end single talk (NearST), far-end single talk (FarST), and double talk (DT). PL: progressive learning. Trans: knowledge transfer. Align: time alignment with MSE loss. VadMask/Vad: VAD with and without masking. Exp 6 is the best configuration of our E2E-AEC.}  
  \begin{tabular}{llcccccccccccc}
    \toprule  
    \multirow{2}{*}{Exp} & \multirow{2}{*}{Method} & \multicolumn{6}{c}{AEC Challenge 2023} & \multicolumn{6}{c}{AEC Challenge 2022} \\
    \cmidrule(lr){3-8} \cmidrule(lr){9-14}
    & & \multicolumn{3}{c}{DT} & \multicolumn{2}{c}{FarST} & \multirow{2}{*}{MOS$_{\text{avg}}$} & \multicolumn{3}{c}{DT} & \multicolumn{2}{c}{FarST} & \multirow{2}{*}{MOS$_{\text{avg}}$} \\
    & & EMOS & DMOS & ERLE & EMOS & DMOS & & EMOS & DMOS & ERLE & EMOS & DMOS & \\
    \midrule  
    \multirow{3}{*}{-} & DeepVQE (E2E) [9] & 4.62 & 4.02 & 65.7 & 4.61 & 4.36 & 4.40 & - & - & - & - & - & - \\
    & Align-ULCNet (Hybrid) [28] & 4.60 & 3.80 & - & 4.77 & 4.28 & 4.36 & - & - & - & - & - & - \\
    & NCA-CRN (E2E) [8] & - & - & - & - & - & - & - & - & - & - & - & 4.29 \\
    \midrule  
    1 & Base Model (E2E) & 4.41 & 3.85 & 46.59 & 4.68 & 4.29 & 4.31 & 4.54 & 4.08 & 46.82 & 4.66 & 4.01 & 4.32 \\
    2 & +PL & 4.48 & 3.96 & 46.39 & 4.68 & 4.41 & 4.38 & 4.60 & 4.16 & 47.35 & 4.68 & 4.16 & 4.40 \\
    3 & +PL+Trans & 4.56 & 4.07 & 49.04 & 4.70 & 4.44 & 4.44 & 4.66 & 4.23 & 48.51 & 4.66 & 4.20 & 4.44 \\
    4 & +PL+Trans+Align & 4.62 & 4.17 & 50.63 & 4.69 & \textbf{4.45} & 4.48 & 4.68 & 4.28 & 50.43 & 4.66 & \textbf{4.27} & 4.47 \\
    5 & +PL+Trans+Align+Vad & 4.64 & \textbf{4.20} & 52.04 & 4.69 & \textbf{4.45} & 4.50 & \textbf{4.69} & \textbf{4.32} & 50.98 & 4.65 & 4.23 & 4.47 \\
    6 & +PL+Trans+Align+VadMask & \textbf{4.65} & 4.18 & \textbf{78.69} & \textbf{4.77} & 4.42 & \textbf{4.51} & \textbf{4.69} & 4.31 & \textbf{79.02} & \textbf{4.73} & 4.24 & \textbf{4.49} \\
    \bottomrule  
  \end{tabular}
  \label{tab:aec_results}  
\end{table*}

\begin{table}[h]
\begin{center}
\footnotesize       
\captionsetup{labelformat=empty}
\caption{Table 2. Comparison of different alignment methods}
\begin{tabular}{lcccccc} 
\toprule 
 & \multicolumn{2}{c}{DT} & \multicolumn{2}{c}{FarST} & NearST & \multirow{2}{*}{MOS $_{\text {avg }}$} \\
 & EMOS & DMOS & ERLE & EMOS & DMOS &  \\
\midrule 
No Align & 4.56 & 4.07 & 49.04 & $\textbf{4.70}$ & 4.44 & 4.44 \\
Attention & 4.57 & 4.09 & 49.61 & 4.69 & 4.41 & 4.44 \\
MSE & 4.60 & 4.08 & 46.80 & $\textbf{4.70}$ & 4.44 & 4.46 \\
Attention+CE & 4.61 & $\textbf{4.18}$ & 50.37 & 4.67 & $\textbf{4.45}$ & $\textbf{4.48}$ \\
Attention+MSE & $\textbf{4.62}$ & 4.17 & $\textbf{50.63}$ & 4.69 & $\textbf{4.45}$ & $\textbf{4.48}$ \\
\bottomrule 
\end{tabular}
\end{center}
\end{table}

\begin{table}[h]
\begin{center}
\footnotesize       
\captionsetup{labelformat=empty}
\caption{Table 3. Impact of VAD predictions from different layers}
\begin{tabular}{lcccccc} 
\toprule 
 & \multicolumn{2}{c}{DT} & \multicolumn{2}{c}{FarST} & NearST & \multirow{2}{*}{MOS $_{\text {avg }}$} \\
 & EMOS & DMOS & ERLE & EMOS & DMOS &  \\
\midrule 
layer 3 & 4.63 & 4.17 & 70.15 & 4.71 & 4.41 & 4.48 \\
layer 5 & $\textbf{4.65}$ & 4.18 & $\textbf{7 8.69}$ & $\textbf{4.77}$ & $\textbf{4.42}$ & $\textbf{4.51}$ \\
layer 8 & 4.64 & 4.18 & 74.86 & 4.75 & 4.39 & 4.49 \\
layer 10 & $\textbf{4.65}$ & $\textbf{4.19}$ & 66.06 & 4.70 & 4.39 & 4.48 \\
\bottomrule 
\end{tabular}
\end{center}
\vspace{-20pt} 
\end{table}

\section{EXPERIMENTS}
\subsection{Data Preperation}
Clean speech and noise clips are taken from the DNS Challenge dataset [24], and room impulse responses (RIRs) are generated using gpuRIR [25]. Echo data is collected from the far-end single talk training clips of the AEC Challenge 2023 [26]. To improve computational efficiency, all audio data is downsampled from 48 kHz (full-band) to 24 kHz (super-wideband), as the perceptual difference between these two sampling rates is negligible [27]. After being processed by the model, the test data is upsampled back to 48 kHz for evaluation.

\subsection{Evaluation Results}
Our evaluation is conducted on the AEC Challenge 2023/2022 blind test sets using AECMOS [29] and echo return loss enhancement (ERLE). AECMOS includes two scores: EMOS for echo annoyance and DMOS for other degradations. In our results, the $MOS_{avg}$ is the average of all AECMOS scores.

In Table 1, the first three rows present E2E and hybrid methods from prior studies, where DeepVQE is a SOTA method. Starting from our E2E base model (Exp 1), Exp 2\~{}6 are progressively optimized configurations. The PL strategy (Exp 2) raises $MOS_{avg}$ from 4.31 to 4.38 on the AEC Challenge 2023. Knowledge transfer (Exp 3) delivers substantial improvements by incorporating knowledge from an LAECbased hybrid system, while time alignment (Exp 4) achieves significant gains through enhanced time synchronization. VAD training (Exp 5) slightly enhances speech quality for DT subset. When combined with masking process (Exp 6), it significantly boosts ERLE, demonstrating remarkable echo suppression in FarST scenarios. Built upon all optimizations consolidated in Exp 6, our E2E-AEC achieves superior results on the AEC Challenge 2023/2022.

To further analyze the time alignment, we perform experiments on the AEC Challenge 2023 based on Exp 4, as shown in Table 2, Loss functions (MSE/CE) are applied by default to attention weights, but for the MSE-only experiment, an additional FC layer is introduced to transform feature dimensions for alignment prediction. The results show that the performance of the Attention-only experiment is almost on par with the baseline (No Align experiment), likely due to the longterm memory of the RNN in our model, which may implicitly capture partial time alignment, thereby limiting the additional benefits provided by applying attention. In contrast, the MSE-only model achieves measurable gains, demonstrating the contribution of loss-driven alignment. Attention+MSE and Attention+CE yield significant improvements, underscoring the effectiveness of integrating attention with loss functions for alignment. Interestingly, the performance differences caused by MSE and CE are almost negligible.

\begin{figure}[h]
  \centering
  \centerline{\includegraphics[width=7.5cm]{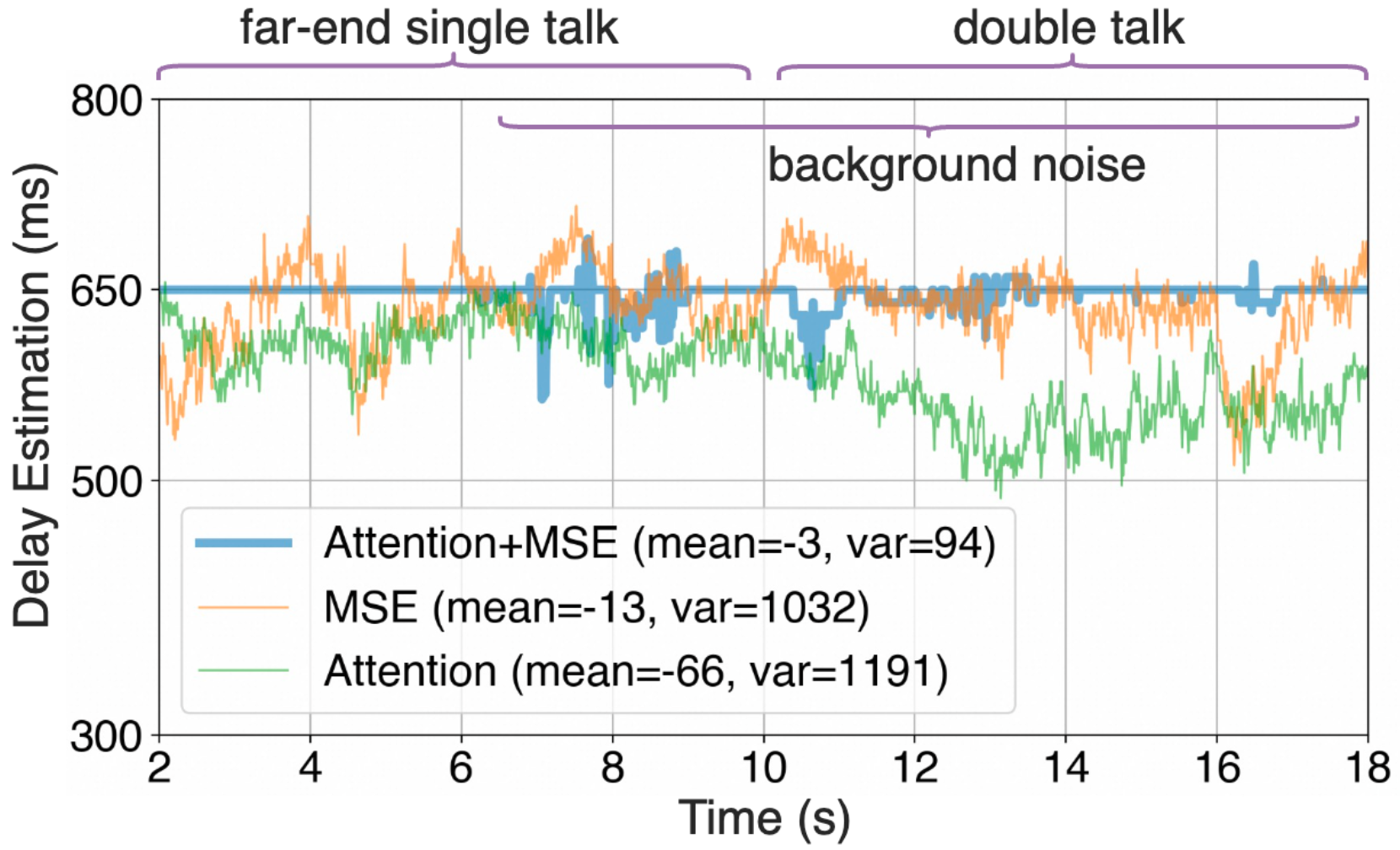}}
\captionsetup{labelformat=empty}
\caption{Figure 2. TDE results for a sample (ground truth delay: 650 ms ). Mean and variance are computed based on the difference between predictions and ground truth.}
\end{figure}

Figure 2 shows TDE results in time alignment, demonstrating the curve for a sample with a ground truth time delay of 650 ms . Due to the convergence time of algorithm, the curve is plotted starting from 2 s. For clarity, Attention+C is omitted, as its performance is similar to Attention + MSE. The figure shows that Attention + MSE achieves the smallest estimation errors, with a mean of only -3 ms and a variance of 94. Minor fluctuations are observed in the Attention+MSE curve near 6 s and 10 s . This behavior matches expectations, as changes in acoustic scenarios involving the introduction of background noise at 6 s and the transition from FarST to DT at 10 s require additional convergence time.

We additionally investigate the impact of VAD prediction and masking from different layers on the AEC Challenge 2023, based on Exp 6, as summarized in Table 3. The fifth layer achieves the best performance, with ERLE reaching 78.69 dB , indicating the highest level of the echo suppression. It suggests that leveraging intermediate layers for VAD prediction and masking is a better strategy.

\section{CONCLUSION}
This paper presents E2E-AEC, an end-to-end neural network model for acoustic echo cancellation, incorporating several key optimizations: progressive learning, knowledge transfer, time alignment, VAD prediction and masking. Experimental results on AEC Challenge datasets demonstrate significant improvements in terms of AECMOS and ERLE, showcasing the effectiveness and robustness of the model across diverse acoustic scenarios. This work underlines the potential of end-to-end methods in addressing the challenges of echo cancellation comprehensively and efficiently.


\clearpage
\section{REFERENCES}
{
\begin{enumerate}[label={[\arabic*]}]
\setlength{\itemsep}{0pt}      
\setlength{\parskip}{4pt}      
\setlength{\parsep}{0pt}       
\item C. Tchassi, "Acoustic echo cancellation for singleand dual-microphone devices: application to mobile devices," in Networking and Internet Architecture, 2013.
\item R. Tyagi, R. Singh, and R. Tiwari, "The performance study of nlmsalgorithm for acoustic echo cancellation," in International Conference on Information, Communication, Instrumentation and Control, 2017, pp. 1-5.
\item G. Enzner, H. Buchner, A. Favrot, and F. Kuech, "Acoustic echo control," Academic press library in signal processing, vol. 4, pp. 807-877, 2014.
\item L. Ma, H. Huang, P. Zhao, and T. Su, "Acoustic echo cancellation by combining adaptive digital filter and recurrent neural network," arXiv preprint arXiv:2005.09237, 2020.
\item R. Peng, L. Cheng, C. Zhang, and X. Li, "Acoustic echo cancellation using deep complex neural network with nonlinear magnitude compression and phase information," in Interspeech, 2021, pp. 4768-4772.
\item Z. Wang, Y. Na, B. Tian, and Q. Fu, "NN3A: neural network supported acoustic echo cancellation, noise suppression and automatic gain control for real-time communications," in ICASSP, 2022, pp. 661-665.
\item G. Zhang, L. Yu, C. Wang, and J. Wei, "Multi-scale temporal frequency convolutional network with axial attention for speech enhancement," in ICASSP, 2022, pp. 9122-9126.
\item Y. Liu, Y. Shi, Y. Li, K. Kalgaonkar, S. Srinivasan, and X. Lei, "SCA: Streaming cross-attention alignment for echo cancellation," in ICASSP, 2023.
\item E. Indenbom, N. Ristea, A. Saabas, T. Parnamaa, J. Guzvin, and R. Cutler, "DeepVQE: Real time deep voice quality enhancement for joint acoustic echo cancellation, noise suppression and dereverberation," in Interspeech, 2023, pp. 3819-3823.
\item Z. Fei and Z. Xueliang, "Attention-enhanced shorttime wiener solution for acoustic echo cancellation," in ICASSP, 2025, pp. 1-5.
\item Y. Jiang and B. Tian, "A small-footprint acoustic echo cancellation solution for mobile full-duplex speech interactions," in ICASSP, 2025.
\item Y. Tu, J. Du, T. Gao, and C. Lee, "A multi-target snr-progressive learning approach to regression based speech enhancement," TASLP, vol. 28, 2020.
\item H. Zhang and D. Wang, "Neural cascade architecture for joint acoustic echo and noise suppression," in ICASSP, 2022, pp. 671-675.
\item L. E. Jeffrey, "Finding structure in time," Cognitive science, vol. 14, no. 2, pp. 179-211, 1990.
\item J. Chung, C. Gulcehre, K. Cho, and Y. Bengio, "Empirical evaluation of gated recurrent neural networks on sequence modeling," in NIPS, 2014.
\item Z. Wang, Z. Wang, and D. Wang, "TF-GridNet: Making time-frequency domain models great again for monaural speech enhancement," in Interspeech, 2022, pp. 376380.
\item Z. Chen, X. Xia, C. Chen, and et al., "A two-stage progressive neural network for acoustic echo cancellation," in Interspeech, 2023, pp. 795-799.
\item G. Hinton, O. Vinyals, and J. Dean, "Distilling the knowledge in a neural network," arXiv, 2015.
\item B. Gholami, M. El-Khamy, and K. Song, "Knowledge distillation for tiny speech enhancement with latent feature augmentation," in Interspeech, 2024, pp. 652-656.
\item C. Knapp and G. Carter, "The generalized correlation method for estimation of time delay," in ICASSP, 1976, pp. 320-327.
\item S. Zhang, Z. Wang, J. Sun, Y. Fu, B. Tian, Q. Fu, and L. Xie, "Multi-task deep residual echo suppression with echo-aware loss," in ICASSP, 2022.
\item T. Vuong, Y. Xia, and R. Stern, "A modulation-domain loss for neural network-based real-time speech enhancement," in ICASSP, 2021.
\item J. Ma and P. Loizou, "SNR loss: a new objective measure for predicting speech intelligibility of noisesuppressed speech," in Speech Communication, 2011, pp. 340-354.
\item H. Dubey, A. Aazami, V. Gopal, and et al., "ICASSP 2023 deep noise suppression challenge," arXiv preprint arXiv:2303.11510, 2023.
\item D. Diaz-Guerra, A. Miguel, and J. Beltran, "gpuRIR: a python library for room impulse response simulation with gpu acceleration," in Multimedia Tools and Applications, 2020.
\item R. Cutler, A. Saabas, T. Parnamaa, and et al., "ICASSP 2023 acoustic echo cancellation challenge," arXiv preprint arXiv:2309.12553v1, 2023.
\item J. Beerends, N. Neumann, E. Broek, A. Casanovas, and et al., "Subjective and objective assessment of full bandwidth speech quality," TASLP, vol. 28, 2020.
\item S. Shetu, N. Desiraju, W. Mack, and E. Habets, "AlignULCNet: Towards low-complexity and robust acoustic echo and noise reduction," 2025.
\item M. Purin, S. Sootla, M. Sponza, A. Saabas, and R. Cutler, "AEC-MOS: A speech quality assessment metric for echo impairment," in ICASSP, 2022, pp. 901-905.
\end{enumerate}
}
\end{document}